\documentclass{article}
\usepackage[left=3cm, right=3cm, top=2cm]{geometry}
\usepackage[utf8]{inputenc}
\usepackage[superscript,nomove]{cite}
\usepackage{xcolor,graphicx}
\usepackage{listings}
\usepackage{listings-golang} 
\usepackage{color}
\usepackage{url}
\usepackage{breakurl}
\usepackage[breaklinks]{hyperref}

\usepackage{cite}

\title{A Blockchain Framework for Managing and Monitoring Data in Multi-Site Clinical Trials}
\author{Olivia Choudhury, Noor Fairoza, Issa Sylla, Amar Das\\
	IBM Research, 75 Binney Street, Cambridge, MA 02142, USA \\
}

\date{}

\begin{document}
	\maketitle

\begin{abstract}

The cost of conducting multi-site clinical trials has significantly increased over time, with site monitoring, data management, and amendments being key drivers. Clinical trial data management approaches typically rely on a central database, and require manual efforts to encode and maintain data capture and reporting requirements. To reduce the administrative burden, time, and effort of ensuring data integrity and privacy in multi-site trials, we propose a novel data management framework based on permissioned blockchain technology. We demonstrate how our framework, which uses smart contracts and private channels, enables confidential data communication, protocol enforcement, and and an automated audit trail. We compare this framework with the traditional data management approach and evaluate its effectiveness in satisfying the major requirements of multi-site clinical trials. We show that our framework ensures enforcement of IRB-related regulatory requirements across multiple sites and stakeholders.

\end{abstract}


\section{INTRODUCTION}

The cost of conducting multi-site research has significantly increased over time, with \$43 billion spent in the US in 2013~\cite{martin2017much, TrialCost}. A significant portion of this cost was incurred by site monitoring, data management, and verification. Typically, a clinical trial involves two phases of amendments to the protocol, each requiring \$450,000. As multi-site trials are expanding to multiple sites to recruit and collect data from more individuals across a diverse population, initiation of such studies experiences a delay due to the sites' reliance on its own institutional review board (IRB) for ethical reviews. To expedite the review process and reduce the administrative burden, time, and cost of redundant reviews, the National Institutes of Health (NIH) issued a policy requiring a single IRB (sIRB) in NIH-funded multi-site clinical research studies~\cite{NIH_sIRB}. Participating sites will still be responsible for obtaining informed consent from subjects prior to enrollment, implementing the approved protocol, and reporting cases of adverse events. Data verification may happen in real time as it is collected, but often occurs through Extract-Transfer-Loads, where problems with data integrity from a site are not detected until later.  Moreover, once a study protocol is approved, it is difficult to track the activities of participating sites and ensure that they adhere to the required guidelines. 

In this context, it is more relevant than ever to design a secure, efficient, and robust infrastructure that not only enforces the regulatory obligations in a multi-site clinical research study, but also ensures an elevated level of data security and optimization of cost. To address the above-mentioned challenges, we propose the design of a novel data management framework based on blockchain technology, smart contracts, and private channels. Blockchain, particularly private blockchain framework, has been proven to be a promising solution for several applications in medical and healthcare research~\cite{choudhury2018enforcing, azaria2016medrec,genestier2017blockchain}. 
Although the authors in~\cite{benchoufi2017blockchain} proposed a public, cryptocurrency-based blockchain for clinical research, it does not provide a practical solution to meet regulatory obligations, where only known parties are allowed to submit study data. Further, in a use case comprising interested stakeholders, the need for cryptocurrencies as incentives is extraneous. Recent literature on auto-generation of smart contracts from clinical trial protocols re-instates the value of blockchain and smart contracts in enforcing protocol guidelines~\cite{CLiPIR,ECliPSE}. 

In this paper, we present a novel, decentralized data management framework based on a private blockchain with private channels to improve data privacy, data integrity, and traceability. We demonstrate how to build a blockchain network comprising different personnel in a multi-site study, create private channels for segregating sensitive data, and leverage smart contracts for enforcing the requirements of a study protocol. We evaluate the effectiveness of our proposed framework in satisfying the major requirements of NIH-funded multi-site clinical trials~\cite{ambrosius2014design,skyler2009intensive,marcus2013randomized,buse2007action}: eligibility criteria for subject enrollment, protection of sensitive data, provenance of data collected from multiple sites, tracking and conducting schedule of activities, timely reporting of adverse events, ensuring valid protocol amendments, audit trail, and assisting in meta-analysis. We compare and contrast our method with the centralized trial management approach. Finally, to provide a holistic overview of implementing blockchain-based solutions for biomedical and healthcare applications, we discuss the challenges entailed by it.

\section*{BACKGROUND}
\subsection{Traditional data management framework for a multi-site clinical trial}
	
In a multi-site clinical trial, the trial protocol must be followed by participating clinics spanning across multiple clinical center networks. Researchers conducting the trial at multiple sites maintain a single database to store all the data related to the trial. Typically, PIs and other clinic staff use a web-based application to access trial information, training material, and upload collected data to a central database, which is located at the coordinating center. 
	
In a traditional clinical trial data management system, when a subject is enrolled, a unique identification number is generated and appended to the corresponding entry on the database. A schedule of activities for each visit is generated by the PI. The web application has a tracking and notification system which can advise clinic staff on subject follow-up windows and required laboratory activities. Tracking a subject begins at screening and continues throughout the trial. During each visit, data is collected in the form of questionnaires, samples, and lab results. The participating clinic sites use the web application to update data from paper-based forms and questionnaires. The data entry system is designed to mirror the forms for accuracy of data entry. This is linked to a relational database that is managed by the coordinating center at a central location. Such a framework is prone to the risk of single point of failure (SPOF) and can result in the loss of data. 
	
\begin{figure*}[!ht]
	\centering
	\includegraphics[scale=0.2]{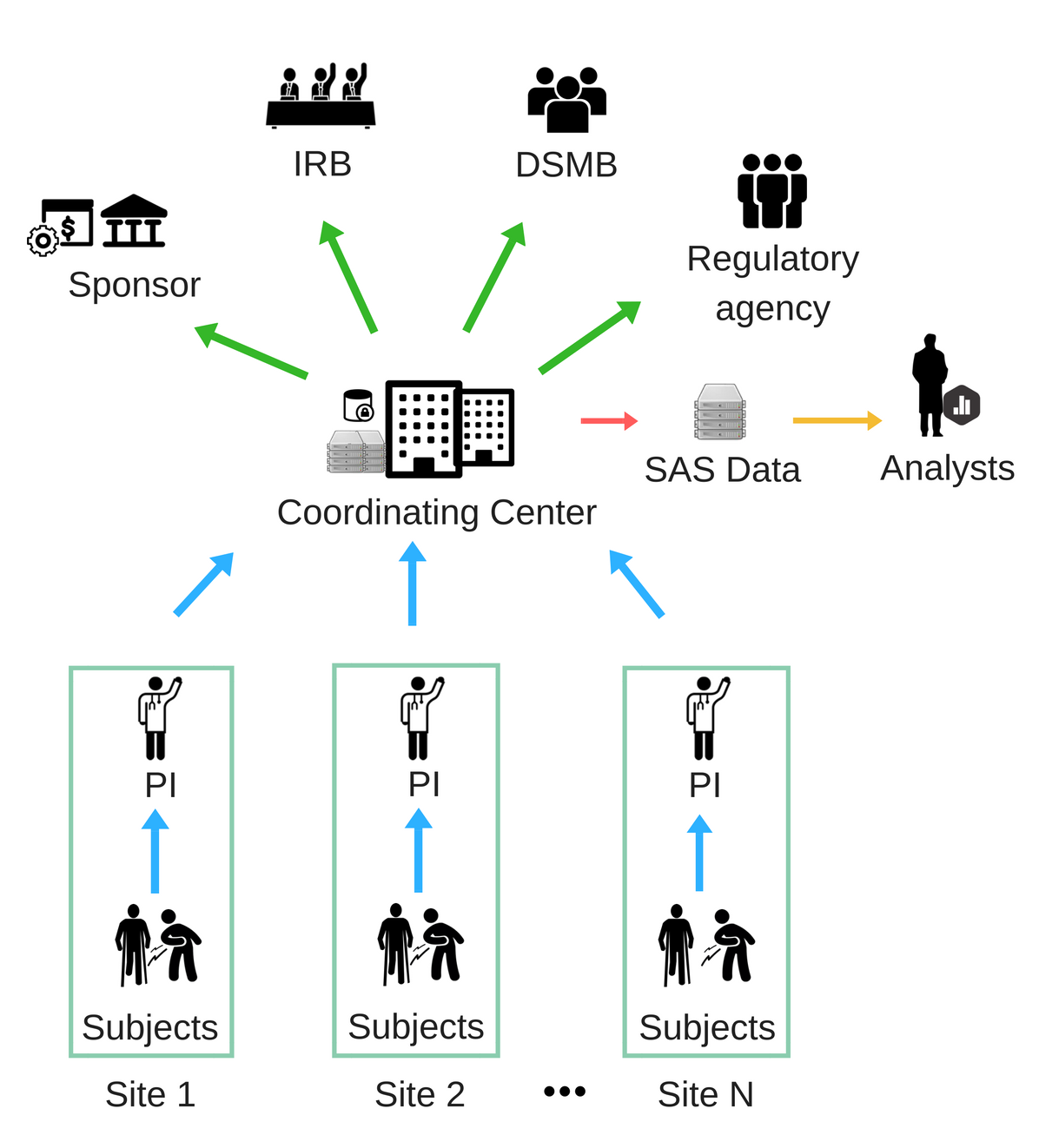}
	\caption{Traditional centralized system of data management and communication in a multi-site clinical trial. Data collected from enrolled subjects at each site is stored in a centralized database maintained by the coordinating center. The database is accessible to the IRB, DSMB, sponsors, and regulatory agency. The de-identified data is converted into SAS format prior to secondary sharing for analysis.} 
	\label{fig:WithoutBC}
\end{figure*}

Following a subject visit, the clinic coordinator reviews the required forms and laboratory reports for accuracy and completeness. 
If the data entered by the clinic staff is overridden, it is flagged for review by the coordinator. The database can also be accessed by the Data and Safety Monitoring Board (DSMB), who review the trial and cases of serious adverse events (SAE) to provide recommendations for continuation of the trial and clinical performance. To assist with trial monitoring and quality control, data edit reports, including site assessments, are generated. The system requires nightly backups and storage of backup tapes in a locked, fire-proof and water-proof area at multiple locations for data recovery, in case of a disaster. Such manual intervention for data management and maintenance increases the cost of trial and the risk of error. Moreover, the data recovery time for such a system is high, which impacts the trial. Figure~\ref{fig:WithoutBC} illustrates the traditional centralized data management framework used in multi-site clinical trials. 

\subsection{Blockchain technology}
Blockchain technology has been perceived to revolutionize several domains, including finance~\cite{nakamoto2008bitcoin, wood2014ethereum}, healthcare~\cite{choudhury2018enforcing, azaria2016medrec, kuo2017blockchain}, supply chain~\cite{kim2016towards}, and the government sector~\cite{BC_Gov_Estonia, BC_Gov_Georgia, BC_Gov_USA}. Blockchain is a decentralized ledger capable of maintaining an immutable record of transactions and tracking assets. The ledger has a verified proof of transactions in blocks that are linked together to form a chain. Each block contains a hash, a list of valid transactions, and a hash of the previous block, attributing to making the blockchain tamper-proof. Although Bitcoin~\cite{nakamoto2008bitcoin} was the first cryptocurrency-based implementation of blockchain, many blockchain frameworks with varying features have been introduced over the last decade. Based on the underlying network, blockchains can be categorized into permissionless and permissioned frameworks. A permissionless blockchain supports anonymous participants to join the network and participate in consensus without any prior approval. They are usually driven by the Proof of Work (PoW) protocol to achieve consensus. Bitcoin and many other cryptocurrency-based implementations are based on such permissionless or public network. Contrary to this, in a permissioned blockchain, only an interested stakeholder with a unique private identity can join the network, read the ledger, propose transactions, and participate in consensus. The transactions can be verified through different consensus mechanisms, such as Proof of Elapsed Time (PoET)~\cite{sawtooth} and Practical Byzantine Fault Tolerance (PBFT)~\cite{fabric}. A majority of permissioned blockchains support smart contracts that define a set of rules to govern transactions within the network. Smart contracts are computer programs that can implement functionalities to enforce rules defined in a business agreement and be executed automatically as part of invoked transactions. They can further enforce authorized data sharing through fine-grained access control~\cite{genestier2017blockchain}.

\subsection{Private blockchains for healthcare applications}

Private blockchain is fundamental in most industry use cases, such as finance, insurance, supply chain, and healthcare. A majority of biomedical and healthcare applications involve interested stakeholders as participants, for which blockchain frameworks built on a private network are more suitable. They do not require incentives, in the form of cryptocurrencies, for participation. Some of the more popular private blockchain implementations include Hyperledger~\cite{hyperledger}, Ethereum~\cite{buterin2014next}, and R3 Corda~\cite{corda}. 
	
In healthcare applications, a private blockchain can be used for secure data collection, management, and sharing. For instance, MedRec~\cite{azaria2016medrec}, a decentralized data management system based on Ethereum's private network, can share electronic medical records between patients and providers. The authors in~\cite{griggs2018healthcare} used Ethereum for secure analysis and management of medical sensors. The Orange Consent Management System~\cite{genestier2017blockchain}, built on Hyperledger Fabric, provides a consent management system for users to share their health data. In the article~\cite{choudhury2018enforcing}, the authors used Hyperledger Fabric to develop a decentralized framework for consent management and secondary use of research data. They also demonstrated how to leverage smart contracts to enforce IRB regulations in a research study.

\subsection{Hyperledger Fabric}
	
\begin{figure*}[!ht]
	\centering
	\includegraphics[width=\textwidth]{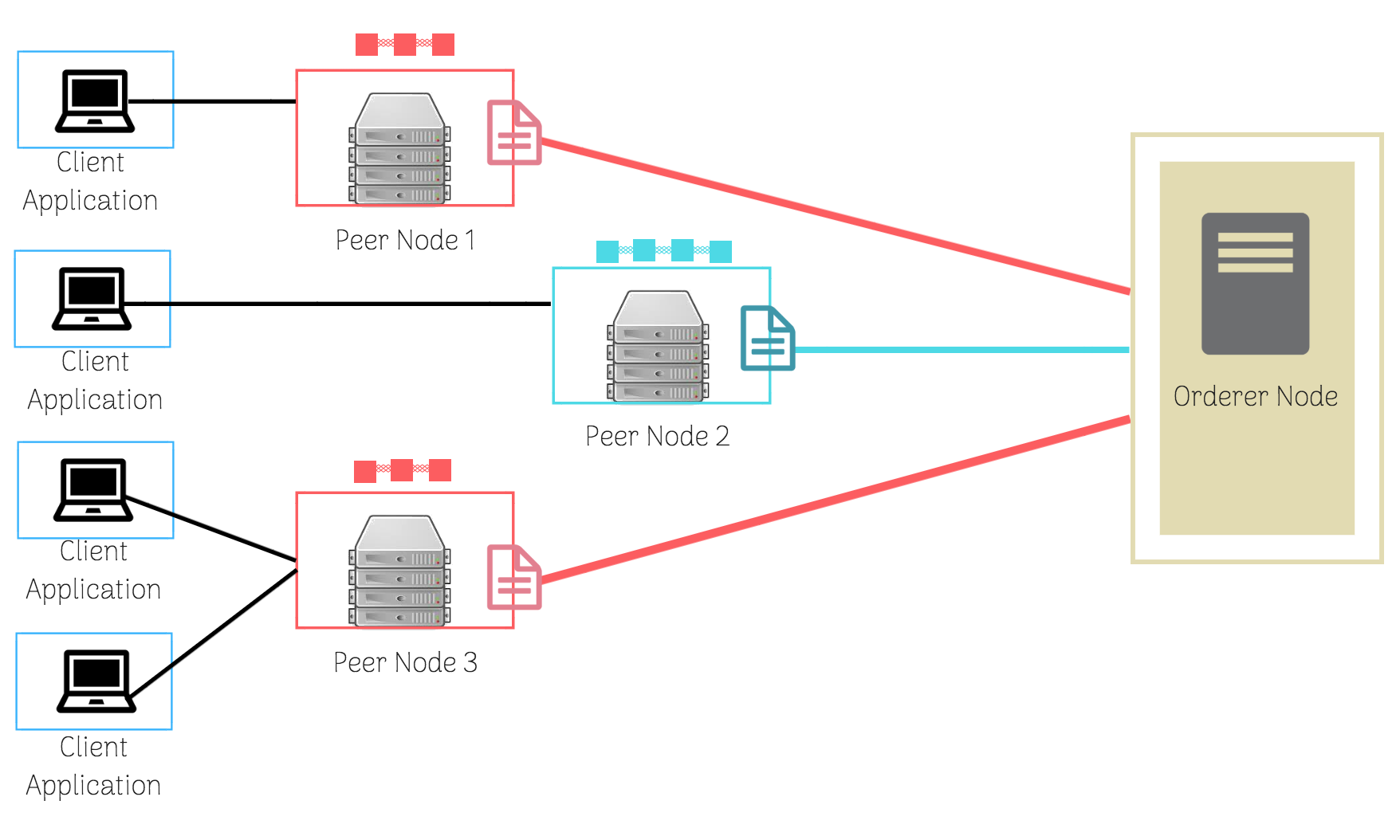}
	\caption{System architecture of Hyperledger Fabric. The private network consists of peer nodes (also acting as client and endorser nodes) and an orderer node. Peer nodes 1 and 3 belong to one channel (shown in red), maintains a ledger, and operates on the same smart contract, whereas peer node 2 belong to a different channel (shown in blue) and has a different ledger and smart contract. The orderer node maintains the order of proposed transactions, validates endorsement signatures, and broadcasts messages to peers.} 
	\label{fig:HF}
\end{figure*}

Hyperledger Fabric is a private blockchain framework developed by The Linux Foundation. It offers a modular architecture supporting pluggable components, such as consensus protocol, encryption, identity management, and membership services. The private network comprises multiple nodes, a smart contract or \textit{chaincode}, and a ledger containing a state database and a log of transactions. A node can be maintained by an individual or multiple users. Based on their functionalities, the nodes can be categorized into client (invokes transactions), peer (maintains and updates ledger), and orderer (supports communication and maintains order of transactions). This is illustrated in Figure~\ref{fig:HF}.
	\begin{itemize}
    	\item Client node: It submits transaction-invocation to the endorsers and transaction-proposals to the orderer. It is connected to both peer and orderer nodes. 
		\item Peer node: It commits transactions and maintains the world state of the ledger. It updates the ledger after receiving ordered states from the orderer. A peer node can act as an endorser to sign a proposed transaction before it is sent to the orderer.
		\item Orderer node: It runs a broadcast communication service to guarantee delivery. It delivers transactions to the peer nodes after verifying the endorsement message.
	\end{itemize}
	
	A smart contract, also known as \textit{chaincode} in Hyperledger Fabric, is a self-executing logic that represents agreements or a set of rules which govern transactions in a blockchain network. The rules are implemented as functions in chaincode. All data transactions that require accessing the ledger invoke corresponding chaincode functions. Hyperledger Fabric implements endorsement policies or conditions to validate proposed transactions. Once a transaction is proposed by a client, it must be endorsed by the pre-defined endorsing nodes. The endorsement signatures are collected and sent to the orderer. The orderer verifies the endorsement message, such as valid number of endorser signatures and simulated transaction results, from all endorsers. The collected transactions are then sent as a new block to all peers. A trusted Membership Service Provider (MSP) enrolls the participants in the network. MSP provides a verifiable digital identity to all entities in the blockchain network, such as peers, orderers, and clients. It serves as a trusted authority which abstracts the process of issuing cryptographic certificates and user authentication. Some of the other notable features of Hyperledger Fabric include:
	\begin{itemize}
		\item Confidentiality: In addition to a private network, it allows creation of private channels comprising a sub-group of network participants. All information related to a channel are accessible only to the members of that channel, thereby elevating the level of confidentiality.
		\item Cryptographic identity management: A membership identity service manages user IDs and authenticates all participants on the network. Additional access control can be implemented for specific network operations.
		\item Modular design: It supports a modular architecture that allows different components, such as membership function and consensus, to be plug-and-play. 
	\end{itemize}

\section{METHODS}
	To address the challenges experienced by a traditional data management system in multi-site clinical trials, we design a blockchain-based infrastructure for collecting, storing, managing, and sharing trial data. We model our system on Hyperledger Fabric to leverage its inherent features, such as private network, private channels, and smart contract. In this section, we describe the design of our proposed system: setting up the network, installing private channels, and writing channel-specific smart contracts.
	
	\subsection{Setting Up a Blockchain Network}
	Prior to setting up a blockchain network, it is important to identify the stakeholders or participants and their roles. Based on a comprehensive study of NIH-funded multi-site clinical trial protocols~\cite{ambrosius2014design,skyler2009intensive,marcus2013randomized,buse2007action}, we enlist the following personnel to be most important in conducting a multi-site clinical trial:
	
	\begin{itemize}
		\item Subject - An individual about whom an investigator collects data to conduct a research study. Data can be collected through intervention or interaction with the individual. 
		\item Principal Investigator (PI) - The primary individual responsible for preparing, conducting, and administering the study at each site of a multi-site clinical trial. While maintaining the ethical conduct and supervision of a study, PIs must protect the rights, safety, and welfare of subjects enrolled in the trial. They have access to the data and specimen collected from subjects, results of analysis, cases of adverse events, and study reports.
        \item Coordinating Center (CC) - A center that is responsible for coordinating a multi-site clinical trial. They generate data edit reports, schedule of activities for subjects, and communicate them to appropriate stakeholders. They maintain a central database that also stores SAS data.
		\item Data Safety and Monitoring Board (DSMB) - An independent group of experts that periodically review and evaluate the accumulated study data for participant safety and study progress, make recommendations concerning the continuation, modification, or termination of the trial.
		\item Institutional Review Board (IRB) - A board designated to protect the rights, safety, and wellbeing of human subjects participating in a clinical trial. They review all aspects of a trial, including approval of study material before and during the study. The new policy of NIH requires a single IRB (sIRB) for conducting ethical review of multi-site studies.
		\item Regulatory agencies - Regulatory agencies, such as The Food and Drug Administration (FDA), ensure medical treatments are safe and effective for subjects to use. They perform inspections of study sites to protect the rights of subjects and to verify the quality and integrity of the data.
		\item Sponsors - An individual, institution, or organization that initiate, manage, and fund a clinical trial, but does not actually conduct it.
		\item Analysts - Researchers or statisticians who gather, review, and organize trial results, including de-identified data collected from subjects.
	\end{itemize}
	
	We consider each personnel involved in a multi-site clinical trial to be a node in the blockchain network. A node can represent an individual (PI), an agency (subject agency), or an organization (sIRB). In a standard blockchain network, each node maintains a single ledger and operates on a smart contract. The smart contract enforces pre-approved guidelines of the study protocol, including the method of collecting, storing, and sharing trial data. Raw data can be directly stored on the ledger. However, for large-scale data, such as wearable sensor readings or genomic data, it is advisable to store it off-chain with an encrypted key linking it to the ledger.

\subsection{Creating Private Channels}
	
	Hyperledger Fabric offers an additional level of data privacy by incorporating private channels between subset of members in the network. A network supports multiple channels, each maintaining a separate smart contract and a ledger. It ensures confidentiality by making all data, including transactions, ledger, members, and channel information invisible and inaccessible to any member not explicitly granted access to that channel. These light-weight channels further reduce the storage space and energy consumption required to maintain a blockchain network. Multiple channels improve transaction throughput by allowing transactions to be executed in parallel~\cite{thakkar2018performance}. For a multi-site clinical trial, we incorporate private channels to streamline data flow and limit data access to designated personnel. We partition the network into the following private channels:

\begin{figure*}[!ht]
	\centering
	\includegraphics[scale=0.3]{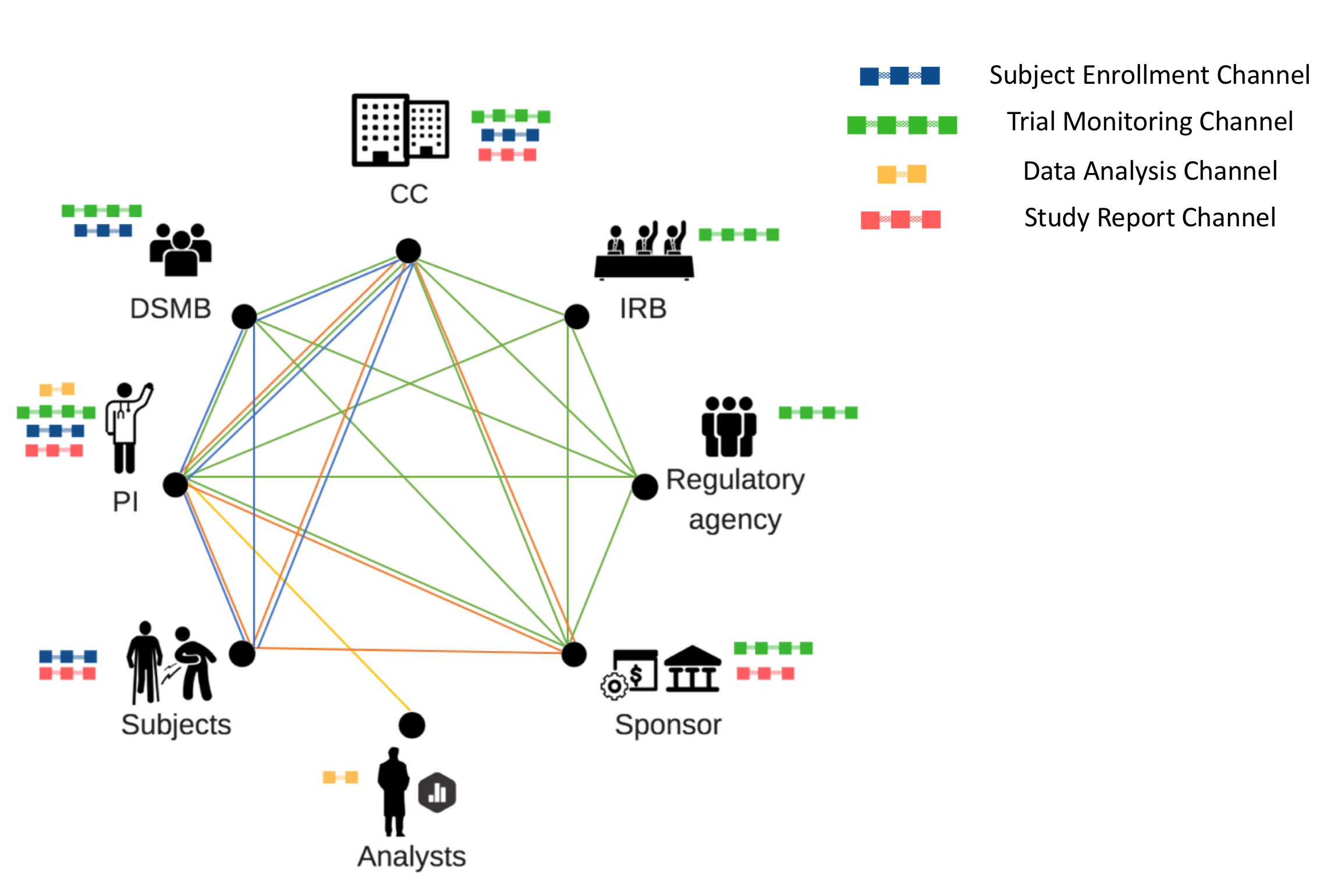}
	\caption{A blockchain-based system with private channels for managing data in a multi-site clinical trial. Each participant maintains a ledger and operates on a smart contract of its member channel. It maintains confidentiality of information and restricts access to channel 
further restrict unauthorized data access by limiting data transactions to channel members.} 
	\label{fig:BC_WithPC}
\end{figure*}

\subsubsection{Subject Enrollment Channel}
	Once subjects are deemed eligible for participation in a clinical trial, PIs collect their data, including consent and protected health information (PHI), for enrollment. This channel corresponds to the consent management system in a clinical trial. To ensure only intended members in the blockchain network have access to such sensitive data, we create a private channel for managing enrollment-specific information. It involves PIs, subjects, coordinating center, and DSMB. This is denoted by the blue channel in Figure~\ref{fig:BC_WithPC}. 
	
	\subsubsection{Trial Monitoring Channel}
	As subjects consent to participate in the study, raw data or samples are collected from them at regular intervals by the trial sites or clinics. We create a separate channel, involving PIs, DSMB, sponsors, coordinating center, and regulatory agencies, for monitoring the activities conducted in a trial. The members of this channel have access to the study data gathered from subjects, activities conducted on each visit, test results, and cases of adverse events experienced during the trial. This is shown by the green channel in Figure~\ref{fig:BC_WithPC}.
	
	\subsubsection{Data Analysis Channel}
	The analysts conduct analyses on de-identified data to get an insight into the effectiveness of the study. We create a data analysis channel between PIs and analysts to enable secondary sharing of de-identified data for analysis. The yellow channel in Figure~\ref{fig:BC_WithPC} depicts this.
	
	\subsubsection{Study Report Channel}
	A clinical study report is a detailed document describing the methods and results of a trial. It integrates clinical and statistical description, presentations, and analyses results. A private channel, comprising PIs, sponsors, regulatory agencies, coordinating center, and subjects, will help in managing and sharing study report documents. This is indicated by the pink channel in Figure~\ref{fig:BC_WithPC}.
	
	\begin{flushleft}
		Hence, in this framework, each node maintains a ledger corresponding to each of its member channel, rather than a single ledger containing the entire trial data. Transactions within each channel is further governed by its own smart contract. Although our proposed system design is based on NIH-funded multi-site clinical trial protocols, it can be extended to meet specific requirements of other studies. 
	\end{flushleft}

\subsection{Implementing Channel-Specific Smart Contracts}
Smart contracts are essential in implementing and enforcing the guidelines of a clinical trial protocol~\cite{choudhury2018enforcing}. These programs can define a list of functionalities incorporating specific requirements for conducting a clinical trial. Blockchain participants can interact with an application interface to invoke the functions in the smart contract. For each data transaction request, the corresponding function verifies if it was proposed by a valid user for a valid channel, data type, and time, thereby allowing fine-grained access control. Herein, we provide a detailed description of the smart contracts specific to the above-mentioned private channels.
	
\subsubsection{Smart Contract for Subject Enrollment}
	Each clinical trial protocol defines a set of inclusion and exclusion criteria that determine if a candidate is eligible to participate in the study. This often requires verifying the age, gender, pre-conditions, and past treatment of the candidates. An enrollment function within the smart contract can check for a set of constraints prior to enrolling candidates and adding their data to the ledger. Since PIs are responsible for de-identification of data, they can trigger a smart contract functionality that will disassociate clinical information from a subject's identifiable information. While enrolling, all the sensitive information collected from the subjects, such as PHI and consent information, can be stored on the ledger. The smart contract can further check if intended members attempt to invoke specific functions. For instance, subjects can attempt to read their data stored by the PIs on the ledger. Similarly, members of DSMB can have read or write access to this ledger. Limiting access of such sensitive information to designated members in the network further elevates data privacy. Once approved, the subject's record will be added to the ledger.

\subsubsection{Smart Contract for Trial Monitoring}\label{SC_for_Monitoring}
	The trial monitoring channel collects and manages different types of data during the course of the trial. In a multi-site research study, each clinic conducts a given set of activities based on a schedule of activities defined in the protocol. This may include collecting questionnaire or samples from subjects, running laboratory tests, and reporting cases of adverse events. A smart contract specific to this channel can verify if the data corresponding to scheduled activities was entered into the system during an approved timeline. It can further check if specific pre-requirements were met prior to data entry. Since current clinical trial management systems often experience difficulty in reporting cases of adverse events or laboratory abnormalities in a timely manner, smart contracts can be leveraged to record, track, and update such events in real-time. In addition to expedited intervention, it provides a transparent and reliable event monitoring approach for the regulatory agency. Moreover, in the case of any discrepancy or attempts to tamper with the recorded data, the smart contract will flag the transaction, which is recorded in the immutable transaction log. In a multi-site trial, this framework ensures provenance and quality of data collected from different sites. The implementation of a single smart contract across all sites further guarantees compliance with IRB-approved study protocol and consistent data framework.

\subsubsection{Smart Contract for Data Analysis}
	The de-identified data collected from subjects are shared with analysts for secondary data analysis. The above-mentioned consent management system coupled with smart contract limit unauthorized data access, even for the least restrictive blanket consent. The smart contract checks the consent information stored on the ledger to ensure the validity of the request. Once approved, de-identified data in SAS format is shared with them. Secondary research includes topics like longitudinal data analysis, subgroup analysis, non-linear relationships, and meta-analysis. Based on its underlying programming language, a smart contract can further invoke modules or functionalities that automatically execute the analysis code.

\subsubsection{Smart Contract for Study Report}
	A clinical study report is an integrated report containing clinical and statistical description, presentations, and analyses. It also includes sample case report forms, information related to the investigatory products, technical statistical documentation, patient data listings, and technical statistical details such as derivations, computations, analyses, and computer output~\cite{food1996guideline}. The smart contract designed for this channel accesses the transaction log to automatically generate audits for evaluating participating sites' performance, protocol adherence, and efficacy of the trial. This can significantly reduce the burden of manual audit and incidents of selective-reporting, under-reporting, or mis-reporting. This is also applicable to periodic generation of reports for continuous reviews. This improves the credibility and integrity of the study report submitted to the regulatory authority.

\begin{flushleft}
Protocol amendments occur frequently in clinical trials, with an average of 2.3 amendments per trial~\cite{getz2008assessing}. When amendments are unavoidable, it is important to execute and track the amendments across all participating sites and stakeholders. Implementing such amendments in our proposed framework would require updating the channel-specific smart contracts such that all intended stakeholders are aware of and comply with the update. This is achieved by a specialized higher level \textit{system chaincode}, that incorporates and tracks updates made to the channel-specific chaincodes. For transparency, prior to initiating an update for a smart contract, an endorsement policy is designed to validate it.
\end{flushleft}

	\section{EVALUATION}
We evaluate the effectiveness of our proposed data management system in enforcing the major requirements of NIH-funded multi-site clinical trials.

\begin{enumerate}
\item \textbf{Audit Trail}: 
As noted by the National Institute of Drug Abuse (NIDA) in its guide to conducting high quality research in the Clinical Trials Network (CTN), proper documentation of how study procedures are implemented is necessary at all times~\cite{NIDA_CTN}. Maintaining audit trail involves regularly monitoring study sites, verifying study compliance, and reporting on monitoring visits. Since current systems rely on manual effort to record and maintain documentation, they experience a delay during staff turnover or unplanned emergencies. Blockchain maintains an immutable log of data transactions and recorded activities, hence eliminating the need for paper-based trails. Any attempts to tamper with the cryptographically secure transactions are automatically flagged and reported. This strict standard for study oversight improves traceability and fidelity of the research. \\

\item \textbf{Data Privacy}: 
Contrary to a traditional system, a private blockchain framework with private channels offers elevated data privacy. As discussed in~\cite{choudhury2018enforcing}, blockchain and smart contracts enable secure consent management and secondary data sharing. PIs can use the system to anonymize data, store sensitive information on the ledger and de-identified data on a database using encrypted keys. Based on the consent information saved on the ledger, third-party researchers access the de-identified data using one-time encryption keys. This restricts them from accessing the data if permissions are revoked. Smart contracts further limit unauthorized data usage through fine-grained access control. Permission for data access can be granted to an individual for a specific data type and duration. Private channels bolster data privacy without affecting transaction efficiency. A recent study showed how a multi-channel setup improves transaction throughput by allowing multiple independent transactions to be executed in parallel~\cite{thakkar2018performance}. \\

\item \textbf{Data Integrity}:
Data provenance and integrity ensure the validity of a study and its findings. Data collection and reporting must follow an approved guideline and standardized representation. This is particularly relevant in multi-site studies, where raw data, questionnaires, and lab reports are collected and curated by multiple participating sites and their personnel. A decentralized system like blockchain expedites data entry, which are time-stamped and nearly impossible to forge once it resides on the system. This automatically enhances trust in the system even in the absence of a centralized trusted authority. To enforce data integrity and coherence across all sites, smart contract functions define and verify a set of criteria that must be met prior to data entry. It can check if eligibility criteria, such as inclusion and exclusion criteria, are met prior to data entry, sites followed a schedule of activities on each visit, and data standards like HL7 and SNOMED CT are followed. This reduces cases of data fraud or inconsistency that are usually detected later in the trial. \\

\item \textbf{Protocol Violation}: 
Protocol violations decrease benefit, increase risk, and affect safety and welfare of the subjects and integrity of the data~\cite{bhatt2012protocol}. Once a study protocol is approved, using the current trial management system, it is difficult to ensure all sites and personnel adhere to it. A smart contract can verify if certain constraints, as required by the protocol, have been met. This includes, but is not limited to, inadequate informed consent, subject enrollment criteria, and mis-reporting or under-reporting of events, to name a few. Through constraint checks and automated monitoring, smart contract functions detect and log cases of protocol violation. It further assists in auto-generating reports to notify regulatory agencies, such as DSMB and IRB, of such incidents. Correct and timely reporting of protocol violations is essential for the interpretation of results and design of post-approval safety assessments of new interventions~\cite{sweetman2011failure}. \\

\item \textbf{Protocol Amendment}:
Amendments to a clinical trial protocol can be challenging for data analysis and interpretation, specially if they occur part way through the trial~\cite{losch2008statistical} and can introduce bias if the changes are made based on the trial data~\cite{rising2008reporting, dwan2011comparison}. The implementation and communication of amendments are also burdensome and potentially costly~\cite{getz2011measuring}. Once a protocol is amended, all sites must adopt and comply with the updated guidelines. Smart contracts, implementing protocol guidelines, are updated upon approval from appropriate members, and installed across all nodes to enforce the updated guidelines and ensure consistency. If an amendment requires updating a channel-specific smart contract, all the other channels can still operate on their existing version. Hence, our framework not only provides a transparent approach of version control, but also a modular design for implementing amendments. \\

\item \textbf{Reporting Adverse Events}: 
To fulfill its obligations during the conduct of a clinical study, an IRB must have information concerning unanticipated problems involving risk to human subjects in the study, including adverse events that are considered unanticipated problems~\cite{FDA_AE}. Failure to report or delayed reporting of an adverse event can have serious consequences on a subject's safety and welfare, study results, and validity of the study. Reporting a case of adverse event is a multi-step process that is often delayed due to manual intervention and lack of a sophisticated approach of reporting. Smart contract functions, triggered by study coordinators, PIs, or affected subjects, enable real-time, automated reporting of adverse events to expedite intervention. These functions can also verify if follow-up studies for assessment were completed. The transaction log can be accessed to easily track and monitor adverse events. Such an approach enables timely identification of adverse events and responsiveness to patient safety concerns. \\
    
\end{enumerate}

\subsection{Comparison}
We compare and contrast our proposed blockchain-based system using private channels with the traditional centralized data management system in Table~\ref{Table:Comparison}. Since the authors in~\cite{thakkar2018performance} have conducted comprehensive experiments to benchmark and evaluate the performance of Hyperledger Fabric, we refer to this work for relevant quantitative evaluations. 

\begin{table*}[h]
		\centering
		\begin{tabular}{p{2.1cm}| p{4.2cm}| p{5.2cm}}
			\hline
			\textbf{Feature} & \textbf{Traditional System} & \textbf{Blockchain with Private Channels}\\
			\hline
			Data privacy & Data can be accessed by study personnel with login credentials & Data access is restricted via encryption, smart contract, and private channels\\
			\hline
			Data storage & Data stored at sites and a central database & Data of a channel is stored on its member nodes' ledger\\
			\hline
            Data recovery & Difficult due to SPOF and requires daily backups and redundant storage & Easy recovery from a decentralized network, even if part of it is down\\
			\hline
			Cost & Low setup cost and high maintenance cost  & High setup cost and low maintenance cost\\
            \hline
		\end{tabular}
		\caption{Comparison of a traditional system and our proposed blockchain-based framework for clinical trial data management.}
		\label{Table:Comparison}
\end{table*}

\section{DISCUSSION}

Multi-site clinical trials entail several challenges, including rising cost and lack of a secure, systematic approach for data management. To address these challenges, we propose a framework based on a permissioned blockchain that supports private channels and smart contracts. We partition the network into multiple channels based on intended data transactions, such as subject enrollment, trial monitoring, data analysis, and study report. Each channel comprises designated stakeholders as members and a smart contract enforcing data transaction guidelines. Building a blockchain network requires significant technical expertise and a relatively higher initial setup cost. However, it offers significant advantages over a traditional centralized system. We demonstrate how our solution satisfies the major requirements of NIH-funded multi-site clinical trials: audit trail, data privacy, data integrity, detecting protocol violation, protocol amendment, and timely reporting of adverse events. 

Although researchers have advocated blockchain technology for healthcare and biomedical applications, it is still at an early stage of development. Despite the hype revolving around this emerging technology, one must discern the need for blockchain and the selection of a suitable framework prior to its adoption~\cite{wust2017you}. The translation of IRB-approved guidelines to smart contract functionalities can be a barrier for the adoption of this technology in clinical trials. To enable seamless transition of protocol guidelines and reproducibility, our ongoing effort also includes developing a framework that automatically generates smart contracts from clinical trial protocols. We have applied machine learning and natural language processing to extract constraints from schedule of activities tables in clinical trial protocols~\cite{CLiPIR}. Through a human-in-the-loop visual interface, we verified the semantics extracted from the tables, such as time of visit, updates to a visit, and activities to be conducted on the visit. As shown in~\cite{ECliPSE}, we converted the extracted information to smart contract functionalities using domain-specific knowledge, represented as ontologies and semantic rules. For future work, we will extend our blockchain-based trial management system such that regulatory agencies can monitor and retrieve necessary information without the burden of maintaining a node for each clinical trial under surveillance. Finally, we will investigate different incentive mechanisms for active participation of stakeholders, particularly in private blockchain frameworks without cryptocurrencies. Currently, we are prototyping and validating our proposed blockchain-based data management framework for a multi-site drug trial.

\section{ACKNOWLEDGMENT}
We would like to thank Prasanna Rao at IBM for valuable comments and suggestions.

\section{REFERENCES}
\bibliographystyle{ieeetr}
\bibliography{Reference}

\end{document}